\documentclass[usenatbib,fleqn]{mnras}
\usepackage{amsmath,amssymb}
\usepackage{txfonts}
\usepackage[T1]{fontenc}
\usepackage{graphicx}
\usepackage{pdflscape} 
\usepackage{afterpage} 

\graphicspath{{./figs/}}
\let\hat=\widehat

\title[AMI Galactic Plane Survey -- Full data release]{AMI Galactic Plane Survey at 16~GHz: II -- Full data release with extended coverage and improved processing}
   
    \author[Perrott et~al.]{Yvette~C.~Perrott$^{1}$\thanks{Corresponding author: email -- ycp21@mrao.cam.ac.uk}, Anna~M.~M.~Scaife$^{2}$, David~A.~Green$^{1}$, Keith~J.~B.~Grainge$^{2}$, \newauthor
    Natasha~Hurley-Walker$^{3}$, Terry~Z.~Jin$^{1}$, Clare~Rumsey$^{1}$, David~J.~Titterington$^{1}$  \\
 $^1$ Astrophysics Group, Cavendish Laboratory, 19 J.~J.~Thomson Avenue, Cambridge CB3 0HE \\
 $^2$ Jodrell Bank Centre for Astrophysics, Alan Turing Building, School of Physics and Astronomy, University of Manchester, Oxford Road, Manchester, \\ M13 9PL, U.K.\\
 $^3$ International Centre for Radio Astronomy Research, Curtin Institute of Radio Astronomy, 1 Turner Avenue, Technology Park, Bentley, WA 6845, Australia \\
}

\date{Accepted ---; received ---; in original form \today}

\pagerange{\pageref{firstpage}--\pageref{lastpage}}

\pubyear{2015}

\begin{document}
\label{firstpage}

\maketitle

\begin{abstract}
The Arcminute Microkelvin Imager Galactic Plane Survey (AMIGPS) provides mJy-sensitivity, arcminute-resolution interferometric images of the northern Galactic plane at $\approx$\,16\,GHz.  The first data release covered $76^{\circ} \lessapprox \ell \lessapprox 170^{\circ}$ between latitudes of $|b| \lessapprox 5^{\circ}$; here we present a second data release, extending the coverage to $53^{\circ} \lessapprox \ell \lessapprox 193^{\circ}$ and including high-latitude extensions to cover the Taurus and California giant molecular cloud regions, and the recently discovered large supernova remnant G159.6+7.3.  The total coverage is now 1777\,deg$^2$ and the catalogue contains 6509 sources.  We also describe the improvements to the data processing pipeline which improves the positional and flux density accuracies of the survey.
\end{abstract}


\section{Introduction}

In a previous paper, \citet{2013MNRAS.429.3330P}, hereafter Paper I, we presented the first data release (DR-I) of the Arcminute Microkelvin Imager Galactic Plane Survey (AMIGPS), a drift scan survey at frequency $\approx$\,16\,GHz.  This provided the most sensitive centimetre-wave Galactic plane survey of large extent at $\nu > 1.4$\,GHz, covering 868\,deg$^2$ of the northern Galactic plane at $\approx$\,3\,arcmin resolution and having noise levels of $\approx$\,3\,mJy\,beam$^{-1}$ away from bright sources.  The AMIGPS provides a new opportunity to characterise the high-frequency emission of Galactic sources, in particular to identify and investigate the properties of sources which are brighter at 16\,GHz than at the lower frequencies (e.g.\ 1.4\,GHz) more often used for Galactic plane surveys.  These include dense plasmas, such as compact {\sc Hii} regions, and the anomalous microwave emission first identified by CMB experiments \citep{1997ApJ...486L..23L} and now demonstrated to exist in many Galactic objects (e.g.\ \citealt{2011MNRAS.418.1889T}, \citealt{2014A&A...565A.103P}).

The second release (DR-II) of the AMIGPS both extends the coverage of the survey and is produced using an improved data processing pipeline.  Here we briefly describe the improvements to the pipeline and present the new data products.

\section{Observations}

The AMIGPS is observed with the AMI Small Array (SA), a radio interferometer located near Cambridge, UK.  It consists of ten 3.7\,m diameter dishes, with baselines ranging between 5 -- 20\,m.  It operates between $\approx$\,12 -- 18\,GHz, with the passband divided into eight channels of 0.72-GHz bandwidth; the bottom two are discarded due to a combination of low response and the presence of geostationary satellites.  The primary beam at the central frequency of $\approx$\,16\,GHz is $\approx$\,20\,arcmin full width at half maximum (FWHM), and the synthesised beam, which is a measure of the resolution, is $\approx$\,3\,arcmin FWHM.  The telescope is sensitive to angular scales of up to $\approx$\,10\,arcmin.  For further details, see \citet{zwart2008}.

The survey is carried out in drift scan mode, meaning that the telescope is pointed at a fixed hour angle and elevation and observes the sky that drifts past.  In practice, the telescope is driven slightly to maintain a fixed J2000 declination, to enable the reobservation of declination strips when necessary.  Declination strips are observed at 12\,arcmin separation to produce a relatively even noise level across the survey; the theoretical noise varies by $\approx\,3$\% between the centre of a declination strip and the point halfway between two strips.  Dates of the observations range from June 2010 to October 2014; no attempt to account for source variability is made when mapping.

\begin{figure*}
  \begin{center}
    \includegraphics[width=0.49\linewidth]{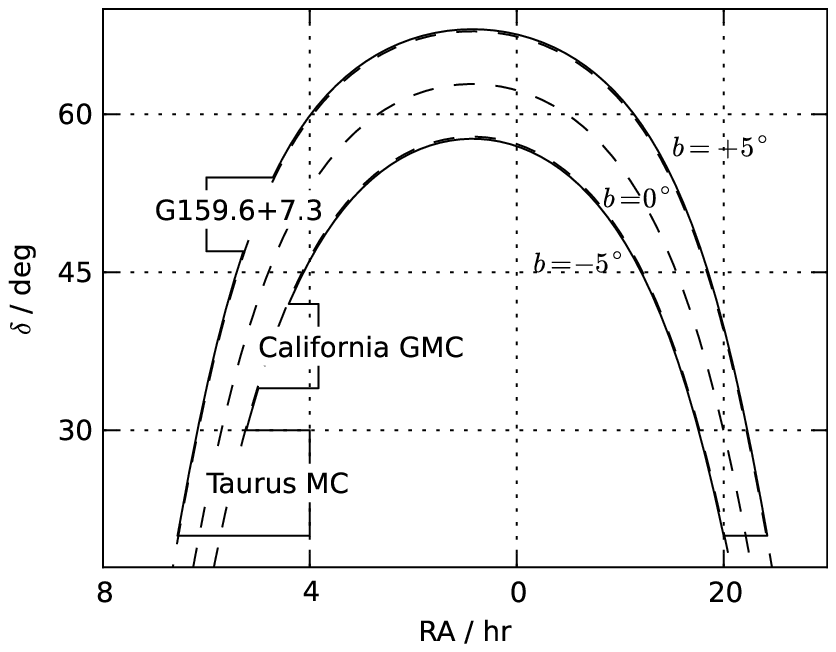}
    \includegraphics[width=0.49\linewidth]{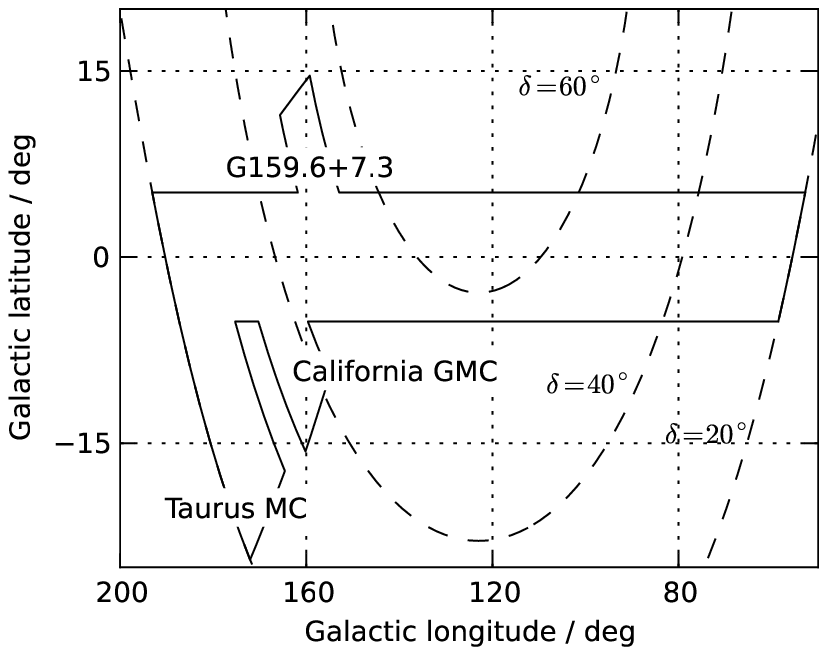}
    \caption{Coverage of the second release of the AMIGPS, in equatorial (left) and Galactic (right) coordinates.}
    \label{Fi:coverage}
  \end{center}
\end{figure*}

The coverage of the survey is shown in Fig.~\ref{Fi:coverage}.  The second data release covers $53^{\circ} \lessapprox \ell \lessapprox 193^{\circ}$, plus high-latitude extensions to cover the Taurus molecular cloud, the California Giant Molecular Cloud \citep{2009ApJ...703...52L}, and the newly discovered large supernova remnant G159.6+7.3 \citep{2010AJ....140.1163F}, for a total of 1777 deg$^2$.

\section{Data reduction and mapping}

Here we briefly outline the data reduction pipeline; for more details see Paper I.  The drift scan data were processed using \textsc{reduce}, an in-house software tool, to flag shadowing, hardware errors and interference, apply phase and amplitude calibrations, and Fourier transform the lag correlator data to synthesise the frequency channels.  Flux density calibration was tied to 3C\,286, based on the \citet{2013ApJS..204...19P} flux density scale.  Discrete pointing centres were defined along each declination strip, and the drift scan data points were associated with and phase rotated to their nearest pointing centre; the data were then output as a multi-source $uv$-\textsc{fits} file.  Data from any reobservations of the same pointing centre were stacked together in $uv$-space.  The pointings were imaged separately in \textsc{aips}\footnote{\textsc{astronomical image processing system} -- www.aips.nrao.edu/}, then corrected for the primary beam attenuation and combined together in a weighted fashion to form a raster map, using another piece of in-house software, \textsc{profile} \citep{2002MNRAS.333..318G}.

Source-finding is carried out at $5\sigma$ on the map, as described in \citet{2011MNRAS.415.2699A}.  The \textsc{aips} task \textsc{jmfit} is used to fit a Gaussian model to each source and the deconvolved source size is used to classify each source as point-like or extended, taking into account the signal-to-noise ratio (SNR) of the source.  See Paper I for more details.

\subsection{Improvements to the pipeline}

This section describes differences to the pipeline described in Paper I, introduced to improve the quality of the final maps.

\subsubsection{Interference flagging}

The interference flagging generally used for AMI data assumes the amplitude of an astronomical signal is constant across the course of an observation, while any spikes are assumed to be caused by interference and are removed.  For drift scan data, this has the unfortunate effect of also removing the peaks of bright sources visible in the timestream data above the noise level; to stop this from occurring, `exclusion zones' were defined in the data where the bright sources are present, and those parts of the timestream were not flagged for interference.  Previously, these exclusion zones were determined iteratively from the $uv$-data, by searching for peaks above the noise level of the data combined over all baselines, however this process occasionally misidentified patches of higher noise and interference as sources.  The exclusion zones are now redefined from the first-pass maps, by searching for sources of peak flux density $\ge 100$\,mJy plus adding some regions of high-surface-brightness extended emission manually, and the pipeline is rerun with the improved exclusion zones.  In addition, a Kolmogorov--Smirnov (KS) test is run on the phase of the data outside the exclusion zones to identify and remove periods of coherent phase caused by interference.  This removes low-amplitude interference that causes striping in the final maps but is not removed by the amplitude interference flagging, present especially at low declination where interference from geostationary satellites is more prevalent.  A small amount of manual interference flagging was also performed to remove remaining stripes visible in the maps.

\subsubsection{Frequency correction}

Since the raster maps are channel-averaged, differences in flagging can introduce variations in the effective central frequency from pointing to pointing.  This is generally a small difference but, particularly at lower declinations where monochromatic interference is often present in the 14.6-GHz channel, can introduce an error in the flux density of up to $\approx$\,5\% for a steep-spectrum ($\alpha = 2$, where $S \propto \nu^{-\alpha}$) source.  To correct for this, we reweighted the channel data for each pointing centre so that the relative weights for each channel are as given in Table~\ref{tab:channel_weights}, giving a constant mean frequency of 15.72\,GHz.  These weights were chosen to represent the natural relative weighting of the channels based on the noise properties of the data, for survey data taken between $40^{\circ} \le \delta \le 45^{\circ}$ which is relatively clean of interference.

\begin{table}
\centering
\caption{Relative weights applied to the survey data, to ensure a consistent central frequency of 15.72\,GHz while preserving as far as possible the natural weighting scheme based on the noise properties of the data.}\label{tab:channel_weights}
\begin{tabular}{lcccccc}\hline
Channel & 3 & 4 & 5 & 6 & 7 & 8 \\\hline
Weight & 0.0908 & 0.223 & 0.204 & 0.200 & 0.182 & 0.0997 \\\hline
\end{tabular}
\end{table}

\subsubsection{Mapping}\label{S:mapping}

The previous \textsc{clean}ing process (see Paper I for details) was found to be over-\textsc{clean}ing in regions around some bright sources, and was modified as follows.  Each pointing was first \textsc{clean}ed to the first negative component, then the thermal noise $\sigma$ on the map was estimated using the \textsc{aips} task \textsc{imean}.  If the brightest pixel was $> 5\sigma$, a circular \textsc{clean} box with radius $\approx\,1.3\times$ the half width at half maximum (HWHM) of the synthesised beam was placed around the pixel, and the pointing was \textsc{clean}ed down to $3\sigma$, then the box was removed.  The pointing was then \textsc{clean}ed down to $3\sigma$ with no boxes.  If the brightest pixel was $\le 5\sigma$, the pointing was \textsc{clean}ed down to $3\sigma$ with no boxes.  In addition, for easier interpretation of the final, combined raster maps, the restoring beam was set as a circular Gaussian with FWHM of 3\,arcmin for all pointings.

\subsubsection{Beam correction}\label{S:beam_corr}

The primary beam for the drift scan data is the weighted average of the primary beams of each of the samples which have been phase rotated to a given pointing centre; it is elongated along the RA axis compared to the standard SA primary beam.  See Paper I for more detail.  In the previous pipeline, the \textsc{clean}ed pointing maps output from \textsc{aips} were multiplied by the inverse of the weighted average primary beam; however, for a source visible in a pointing away from the centre, this had the effect of skewing the restoring Gaussian and shifting its centroid by a small but significant amount.  In the new pipeline, we instead correct the \textsc{clean} components for the weighted average primary beam and then convolve the corrected \textsc{clean} components with the restoring Gaussian, and add the (beam-corrected) residuals to produce the final pointing map.

We have also improved the description of the primary beam since, from inspection of flux densities of bright, point-like sources appearing at various places in individual pointing centre maps, the Gaussian approximation generally used for the AMI primary beam was found to underestimate the beam correction toward the edges of the maps.  When flux densities were measured from the final, combined maps, this caused a small ($\approx$\,1--2\%) underestimation.  To improve the characterisation of the beam toward the edges, we used sets of drift scans around the bright, point-like radio galaxy 4C\,39.25 to fit for the parameters in the (more flexible) \textsc{aips} parameterisation:

\begin{align}\label{eq:aips_pb}
\mathrm{Primary\:beam} = &1 + x \times \mathbf{pbparm(3)}/10^{3} + x^2 \times \mathbf{pbparm(4)}/10^{7} \nonumber \\ 
&+ x^3 \times \mathbf{pbparm(5)}/10^{10} + x^4 \times \mathbf{pbparm(6)}/10^{13} \nonumber \\
&+ x^5 \times \mathbf{pbparm(7)}/10^{16},
\end{align}
where \textbf{pbparm($i$)} are fitting parameters and $x = $ (distance from the pointing centre in arcmin $\times$ frequency in GHz)$^2$.  There is significant degeneracy between the \textbf{pbparm($i$)} parameters, so we constrained each one to lie on a straight line as a function of frequency in Hz, i.e.\ \textbf{pbparm($i$)}$ = m_{i} \times (\nu - \nu_{0}) + $\textbf{pbparm($i$)}$_{0}$ where $\nu_{0}$ is set to 15.75\,GHz.  Allowing \textbf{pbparm(7)} to vary did not result in a significant improvement in the fit, so we fix it to 0.  The best fit parameters are shown in Table~\ref{tab:pbparms}.

\begin{table}
\centering
\caption{Fitted parameters for the \textsc{aips} parameterisation of the SA primary beam as a function of frequency, i.e.\ \textbf{pbparm($i$)}$ = m_{i} \times (\nu - \nu_{0}) + $\textbf{pbparm($i$)}$_{0}$ where $\nu_{0}$ is set to 15.75\,GHz.}\label{tab:pbparms}
\begin{tabular}{l r r} \hline
$i$ & \textbf{pbparm($i$)}$_{0}$ & $m_{i}$ \\ \hline
  3 & $-2.76102\times 10^{-2}$ & $4.50461\times 10^{-13}$ \\
  4 & $3.24852\times 10^{-3}$ & $-1.79711\times 10^{-14}$ \\
  5 & $-1.89970\times 10^{-5}$ & $-2.40098\times 10^{-16}$ \\
  6 & $4.21923\times 10^{-8}$ & $-2.77453\times 10^{-19}$ \\
  7 & $0.00000$ & $0.00000$ \\ \hline
\end{tabular}
\end{table}

\begin{figure*}
  \begin{center}
    \includegraphics[bb=57 332 556 464, clip=, width=\linewidth]{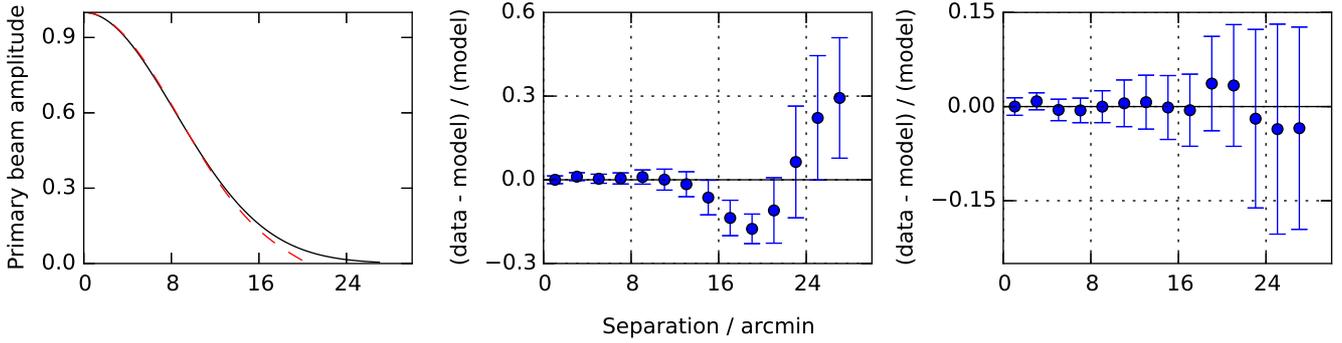}
    \caption{The left-hand plot shows the standard Gaussian primary beam model (black, solid line), and new, fitted \textsc{aips} parameterisation (red, dashed line), for the central frequency.  The two right-hand plots show fractional residuals with respect to the Gaussian primary beam model (centre), and new, fitted \textsc{aips} parameterisation (right) at the central frequency, for a set of channel-averaged drift scans around 4C\,39.25.  The data are averaged in bins of 2\,arcmin in separation from the source, and the errorbars show the standard deviation within each bin.  Note the difference in $y$-axis scale.  In the AMIGPS data reduction pipeline, the weighted, average primary beam is cut off at the 10\% power point.}
    \label{Fi:pb_residuals}
  \end{center}
\end{figure*}

The resulting improvement to the agreement with the data, and a comparison between the two beam parameterisations are shown in Fig.~\ref{Fi:pb_residuals}.  Note that in the data reduction pipeline the weighted, average primary beam is cut off at the 10\% power point.

\subsubsection{Combining maps}

Before combining the individual maps into the large raster maps, the individual maps are regridded to the raster map grid using the \textsc{aips} task \textsc{regrd}, to take advantage of this more sophisticated regridding algorithm compared to the simple algorithm used for the task previously in \textsc{profile}.  This has the effect of increasing source flux densities measured from the combined maps by a mean of $\approx$\,1\%.

\section{Comparison to DR-I}

To investigate the differences between DR-I and DR-II, we matched sources within 1$\,$arcmin ($\approx\,2\sigma$ for the lowest SNR of 5 for the larger beam size of 3$\,$arcmin used in DR-II) in the two catalogues.  This produced 3157 matches, i.e.\ 90\% of the sources in DR-I have matches in DR-II.  The 346 sources without matches in the DR-II catalogue are either low-SNR and have dropped below the 5$\sigma$ threshold due to repositioning of noise spikes; are on or near patches of extended emission and have had their centroid shifted due to the different imaging procedures; or are near a bright source whose exclusion zone has changed and are now excluded.

\begin{figure}
  \begin{center}
    \includegraphics[bb=206 316 400 475, clip=]{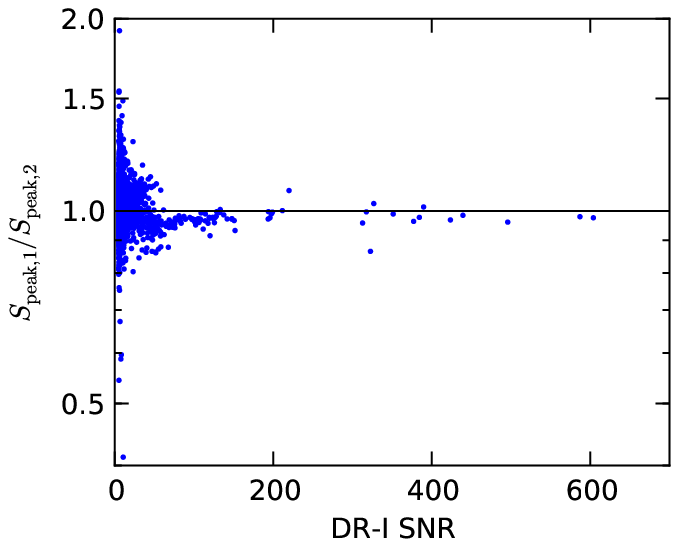}
    \includegraphics[bb=206 316 400 475, clip=]{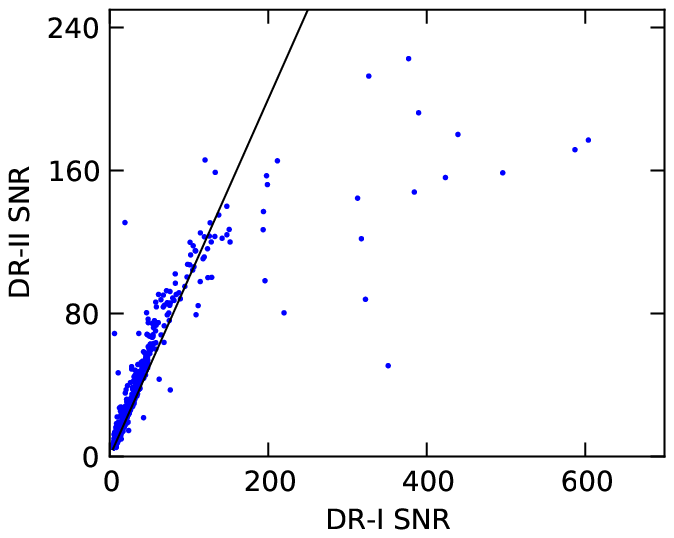}
    \caption{Comparison between peak flux densities (top) and SNRs (bottom) for sources classified as point-like in both DR-I and DR-II.  The black line shows the one-to-one correspondence in both cases.  The improvements in the pipeline have produced a slight increase in flux density in DR-II compared to DR-I.  For low-to-medium-SNR sources, the SNR is increased slightly, while for high-SNR sources it is decreased.}
    \label{Fi:S_comp}
  \end{center}
\end{figure}

2423 of the sources are classified as point-like in both the DR-I and DR-II catalogues.  The peak flux densities and SNRs for these sources are compared in Fig.~\ref{Fi:S_comp}.  It can be seen that flux densities are generally slightly higher in DR-II; this is due to the improved primary beam correction as explained in Section~\ref{S:beam_corr}.  Also, SNRs for low-to-medium-SNR sources are increased slightly due to the improvements in interference flagging; SNRs for high-SNR sources are decreased due to the improvement in the \textsc{clean}ing algorithm which no longer artificially decreases the noise around bright sources.  Comparisons for extended sources are more complicated and are not attempted here.

\section{Calibration accuracy of DR-II}

\subsection{Positional accuracy}

Similarly to Paper I, the errors $\sigma_{\mathrm{RA}}$ and $\sigma_{\delta}$ in RA and $\delta$ for a point source are assumed to be given by 
\begin{equation}
\label{eqn:pos_err}
\sigma_{\mathrm{RA\: or\:} \delta}^2 = \epsilon_{\mathrm{RA\: or\:} \delta}^2 + \frac{\theta^2}{2\ln(2) \: \mathrm{SNR}^2},
\end{equation}
where $\epsilon_{\mathrm{RA\: or\:} \delta}$ are the r.m.s.\ calibration errors in RA and $\delta$ and the second term is the noiselike uncertainty in each direction (now the same in both directions since the beam is round); $\theta$ is the FWHM of the synthesised beam, always 180\,arcsec in DR-II.

For DR-II, we found $\epsilon_{\mathrm{RA}}=1.6$\,arcsec and  $\epsilon_{\delta}=1.9$\,arcsec by minimising the KS statistic for the separations between the detected and true (to milli-arcsec accuracy) positions of 252 Very Long Baseline Array Calibrator Survey (VCS, \citealt{2002ApJS..141...13B}) catalogue sources (c.f.\ 2.6 and 1.7\,arcsec respectively in DR-I).  The two calibration error terms are now very similar, indicating that the systematic effect introduced by primary-beam-correcting the restoring Gaussian (see Section~\ref{S:beam_corr}), which was worse in the RA direction due to the beam elongation, has been eliminated by the improved pipeline.

As in DR-I, the positional uncertainties for extended sources are calculated as:
\begin{equation}
\label{eqn:pos_err_ext}
\sigma_{\mathrm{RA\: or\:} \delta}^2 = \epsilon_{\mathrm{RA\: or\:} \delta}^2 + \sigma_{\mathrm{J,RA\: or\:J,} \delta}^2 ,
\end{equation}
where the appropriate $\sigma_{\mathrm{J}}$ terms are the errors estimated by the \textsc{aips} fitting task \textsc{jmfit}, which folds in an estimate of the noiselike error as well as the error associated with the fit.

\subsection{Flux density accuracy}

As in DR-I, we assume flux calibration errors are given by

\begin{equation}
\label{eqn:flux_err}
\sigma_{S}^2 = (0.05\,S)^2 + \sigma^2,
\end{equation}
where $S$ is peak (integrated) flux density for a point-like (extended) source.  This error estimation comprises a 5\% calibration uncertainty and a noise-like error $\sigma$ which for a point-like source is the r.m.s.\ map noise measured from the \textsc{clean}ed map, and for an extended source is the error estimated by \textsc{jmfit} which also folds in an estimation of the fitting error.  Note that this does not account for the effect of flux loss for extended sources.

\begin{figure}
  \begin{center}
    \includegraphics[bb=180 300 427 492, clip=, width=\linewidth]{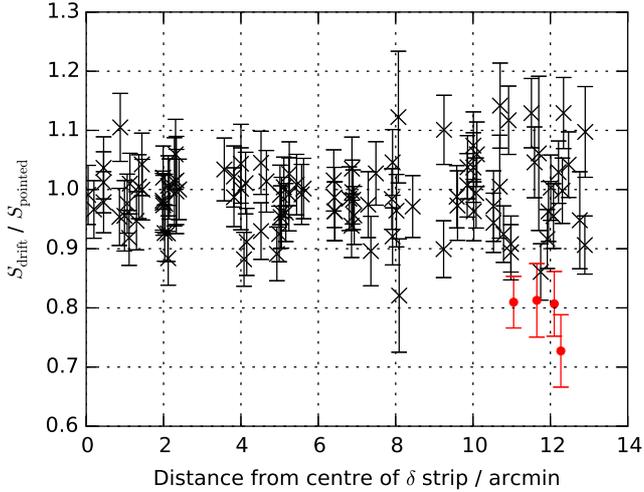}
    \caption{Ratios of flux densities measured from drift scan observations and mean of pointed observations taken within 10 days, as a function of the position of the source with respect to the centre of the drift scan observation.  The errorbar shows the drift scan error only.  Black crosses (red dots) show points at $\le$ ($>$) $3\sigma$.}
    \label{Fi:flux_comp}
  \end{center}
\end{figure}

We tested the flux density calibration accuracy by searching for semi-concurrent pointed observations of VCS sources, which are used as phase calibrators for AMI observations and so are observed on a regular basis.  We remapped individual drift scan observations containing these sources using the standard pipeline and compared the source flux density measured from the drift scan map to the mean of any pointed observation flux density measurements taken within 10 days (to limit variability), for 111 drift scan/pointed observation matches.  Fig.~\ref{Fi:flux_comp} shows the ratio of drift scan to pointed observation flux density as a function of the distance of the source from the centre of the declination strip; the worst outliers lie at the edge of the strip, where direction-dependent effects (which can suppress flux) are expected to be worst.

96\% of the drift scan flux densities lie within $3\sigma$ of the pointed observation flux density (c.f.\ 93\% for DR-I).  Excluding the $3\sigma$ outliers and taking the median percentage difference shows that the drift scan flux densities are biased low by $\approx\,1$\% (c.f.\ 2\% for DR-I); this is a common effect in surveys (see, e.g.\ \citealt{2011MNRAS.415.2708A}) and we do not correct for this small error.  This bias is fairly insensitive to the number of days within which the match is performed, and the exclusion of points at greater distance from the centre of the strip.

\section{Completeness}

Estimation of the completeness of the survey, or the fraction of sources expected to be detected over the survey region at a given flux density level, is made difficult due to the `exclusion zones' which were employed around bright sources to account for elevated, non-Gaussian noise present around sources with peak flux density $\gtrapprox$\,50\,mJy (see Paper I).  Sources within the `exclusion radius' of a bright source with peak flux density $S_{\mathrm{peak,bright}}$ were required to have peak flux density $S_{\mathrm{peak}} \geq S_{\mathrm{peak,bright}}/10$ to be included in the catalogue, effectively redefining the noise level within the exclusion zone to be $\max ( \sigma_{\mathrm{orig}}, S_{\mathrm{peak,bright}}/50)$ where $\sigma_{\mathrm{orig}}$ is the existing thermal noise estimate from the map.

\begin{figure}
    \includegraphics[bb=180 296 427 494, clip=]{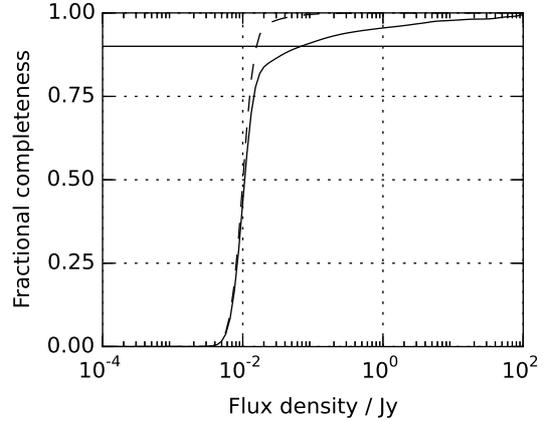}
    \caption{An estimation of the completeness of the AMIGPS DR-II calculated from the noise maps.  The completeness was calculated over the whole survey area (solid line) and outside the exclusion zones around bright sources only (dashed line).  The horizontal line shows the 90\% completeness limit.}
    \label{Fi:completeness}
\end{figure}

The probability of a source with true flux density $\hat{S}$ being detected when lying on a pixel with thermal noise $\sigma_{\mathrm{n}}$ is given by

\begin{eqnarray}
\label{eqn:completeness}
P(\hat{S} \geq 5 \sigma_{\rm{n}}) = \int^{\infty}_{5\sigma_{\rm{n}}} \frac{1}{\sqrt{2\pi\sigma_{\rm{n}}^{2}}}\exp{\left(-\frac{\left(x - \hat{S}\,\right)^{2}}{2\sigma_{\rm{n}}^{2}}\right)}~\mathrm{d}x,
\end{eqnarray}
assuming Gaussian statistics.  The theoretical probability of the source being detected can therefore be calculated by averaging the probabilities given by Eqn.~\ref{eqn:completeness} for each pixel in the map.  This is illustrated in Fig.~\ref{Fi:completeness} and was calculated by (a) using pixels outside the exclusion zones only (dashed line) and (b) using all pixels in the map, and assuming that the effective noise level defined above for the pixels inside exclusion zones can be treated as an approximation to the Gaussian noise level $\sigma_{\mathrm{n}}$ in Eqn.~\ref{eqn:completeness} (solid line).

The accuracy of these completeness curves was also tested via simulation.  Simulated sources at a given flux density were inserted in the maps using the \textsc{aips} task \textsc{immod} at random positions; the standard source-finding pipeline was run on the maps and the fraction of sources detected was recorded for each peak flux density.  The results of the simulations agreed very well with the theoretical curves in Fig.~\ref{Fi:completeness}.

Outside the exclusion zones, the survey is 90\% complete above $\approx$\,16\,mJy, but when including the exclusion zones it does not reach 90\% completeness until $\approx$\,68\,mJy.  It should be noted however that any effect due to correlation between source positions is not included in the completeness estimation.  The completeness curve for the whole survey is expected to be slightly overestimated due to this effect.

\section{Data products}\label{S:data_products}

The maps and full source catalogue are now available online, at \url{http://www.astro.phy.cam.ac.uk/surveys/amigps} and from CDS.  Maps are presented in Galactic coordinates covering the main survey area and the extensions separately.  Each map also has an associated noise map and `adjusted' noise map which shows the effective noise level after adjusting for bright sources (see Paper I for more details).  We show two example maps in Section~\ref{S:maps}.

The catalogue contains a total of 6509 sources.  As in DR-I, for each source the catalogue contains:

\begin{itemize}
\item{A source name, constructed from the J2000 RA and $\delta$ coordinates of the source.}
\item{The peak RA (RA$_{\mathrm{peak}}$), $\delta$ ($\delta_{\mathrm{peak}}$), flux density ($S_{\mathrm{peak}}$) and associated errors ($\Delta \mathrm{RA}_{\mathrm{peak}}$, $\Delta \delta_{\mathrm{peak}}$, $\Delta S_{\mathrm{peak}}$); these are the appropriate quantities to use for point-like sources.}
\item{The fitted centroid RA (RA$_{\mathrm{cent}}$) and $\delta$ ($\delta_{\mathrm{cent}}$), integrated flux density ($S_{\mathrm{int}}$) and associated errors ($\Delta \mathrm{RA}_{\mathrm{cent}}$, $\Delta \delta_{\mathrm{cent}}$, $\Delta S_{\mathrm{int}}$); these are the appropriate quantities to use for extended sources.}
\item{The critical source size ($e_{\mathrm{crit}}$, size below which a source is classified as point-like; see Paper I) and the deconvolved source major and minor axis sizes and position angle ($e_{\mathrm{maj}}$, $e_{\mathrm{min}}$, $e_{\theta}$).  A deconvolved size of 0.0 indicates that the source was not found to be wider than the synthesised beam in the major or minor axis direction.}
\item{The $\chi^2$ value for the fit (note that this is an indication of the goodness of fit of the Gaussian model to the source, not of the reliability of detection of the source).}
\item{The source classification (point-like or extended).}
\end{itemize}

\noindent A sample section of the catalogue is shown in Table~\ref{T:examp_cat}.

\section{Example maps and catalogue}\label{S:maps}

\subsection{G90.70+2.20}

\begin{figure}
    \includegraphics[bb=46 113 567 664, clip=, width=\linewidth]{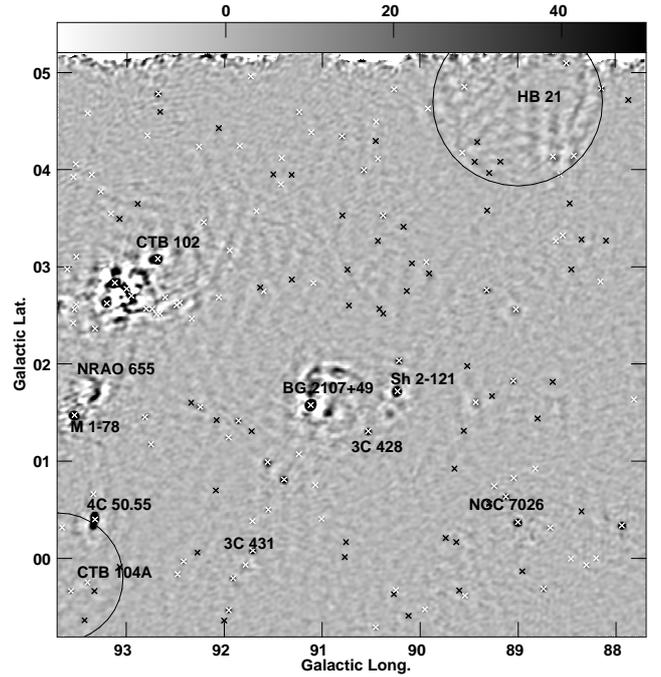}
    \caption{One of the AMIGPS $6^{\circ}\times 6^{\circ}$ square raster maps covering the main survey area.  The grey-scale is in mJy\,beam$^{-1}$ and has been truncated at $-20$ and $+50$ to show low-surface-brightness features.  Detected sources are marked with $\times$, and several known objects are marked (see text for details).  The large circles mark the approximate outlines of the supernova remnants HB\,21 and CTB\,104A.  The resolution of the map is 3\,arcmin.}
    \label{Fi:G90}
\end{figure}

\begin{figure}
    \includegraphics[bb=46 113 567 664, clip=, width=\linewidth]{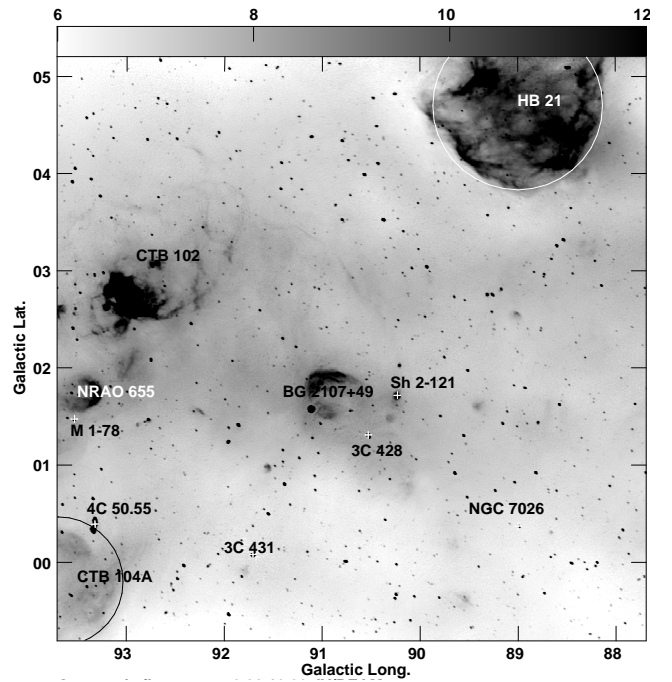}
    \caption{The same area as Fig.~\ref{Fi:G90}, as seen in the CGPS.  The grey-scale shows brightness temperature in K and has been truncated at 6 and 12 to show low-surface-brightness features.  Several known objects are marked (see text for details).  The large circles mark the approximate outlines of the supernova remnants HB\,21 and CTB\,104A.  The resolution of the map is $\approx\,1.3$\,arcmin.}
    \label{Fi:CGPS}
\end{figure}

Fig.~\ref{Fi:G90} shows one of the $6^{\circ}$ square raster maps covering the main survey area, centred at $\ell = 90\fdg 7$, $b = +2\fdg 20$.  There are 160 sources detected in this map, of which 120 are classed as point-like, 32 as extended, and 8 as undefined (the Gaussian fit failed in these cases).  For comparison in Fig.~\ref{Fi:CGPS} we show the (total-power) Canadian Galactic Plane Survey (CGPS; \citealt{2003AJ....125.3145T}) map at 1.4\,GHz covering the same region.  As an example of the AMIGPS catalogue format we show the catalogue entries for the brightest 20 sources in this field in Table~\ref{T:examp_cat}.  For ease of comparison with Fig.~\ref{Fi:G90} we also give Galactic coordinates in Table~\ref{T:examp_cat}; these are not included in the AMIGPS catalogue.

The AMI map shows features covering a range of angular scales, from the very large, such as the supernova remnants HB\,21 \citep{1953MNRAS.113..123H} and CTB\,104A \citep{1963AJ.....68..181W} and the \textsc{Hii} region complexes CTB\,102 \citep{1960PASP...72..331W}, BG\,2107+49 \citep{1984JRASC..78..155H}, and NRAO\,655 \citep{1966ApJS...13...65P}, to the medium, such as the \textsc{Hii} region Sh\,2-121 \citep{1959ApJS....4..257S} which is extended slightly to the AMIGPS beam, to the point-like, such as the planetary nebula NGC\,7026 \citep{1888MmRAS..49....1D} and the compact \textsc{Hii} region M\,1-78 \citep{1946PASP...58..305M}.  Some extra-galactic sources are also visible, including 3C\,428, 3C\,431 \citep{1959MmRAS..68...37E} and 4C\,50.55 \citep{1965MmRAS..69..183P}.  From comparison to the CGPS map it is clear that only the sharp knots and filaments in the extended structures are visible, and the smooth, large-scale emission is resolved out.

\afterpage{
\begin{landscape}
\centering
\begin{table}\scriptsize
\caption{An example catalogue showing the parameters of the brightest 20 sources detected in Fig.~\ref{Fi:G90}.  See Section~\ref{S:data_products} for a description of the quantities.  Coordinates are J2000, coordinate errors are in arcsec, flux densities are in mJy, and angular sizes are in arcsec.  Galactic coordinates (in degrees) are also added for ease of comparison with Fig.~\ref{Fi:G90}; these are not in the AMIGPS catalogue.}\label{T:examp_cat}
\begin{tabular}{lrrrrrrrrrrrrrrrrrrrr}
\hline
Source & RA$_{\mathrm{peak}}$ & $\delta_{\mathrm{peak}}$ & $\Delta \mathrm{RA}_{\mathrm{peak}}$ & $\Delta \delta_{\mathrm{peak}}$ & $\ell_{\mathrm{peak}}$ & $b_{\mathrm{peak}}$ & $S_{\mathrm{peak}}$ & $\Delta S_{\mathrm{peak}}$ & RA$_{\mathrm{cent}}$ & $\delta_{\mathrm{cent}}$ & $\Delta \mathrm{RA}_{\mathrm{cent}}$ & $\Delta \delta_{\mathrm{cent}}$ & $S_{\mathrm{int}}$ & $\Delta S_{\mathrm{int}}$ & $e_{\mathrm{crit}}$ & $e_{\mathrm{maj}}$ & $e_{\mathrm{min}}$ & $e_{\theta}$ & $\chi^2$ & Type \\ \hline
AMIGPS J210937+501334 & 21:09:37.21 & +50:13:34.1 & 2.6 & 2.8 & 91.12 & 1.58 & 1073.9 & 55.7 & 21:09:36.40 & +50:13:36.6 & 2.0 & 2.2 & 1425.2 & 77.4 & 100.0 & 114.9 & 81.7 & 106.6 & 2.91 & E \\
AMIGPS J212044+515324 & 21:20:44.85 & +51:53:24.3 & 2.2 & 2.5 & 93.53 & 1.47 & 901.6 & 46.0 & 21:20:44.62 & +51:53:23.4 & 1.8 & 2.1 & 857.3 & 45.5 & 100.0 & 16.2 & 0.0 & 21.2 & 1.59 & P \\
AMIGPS J211229+523240 & 21:12:29.71 & +52:32:40.0 & 6.4 & 6.5 & 93.12 & 2.84 & 630.9 & 40.7 & -- & -- & -- & -- & -- & -- & 109.1 & -- & -- & -- & -- & -- \\
AMIGPS J210920+522333 & 21:09:20.90 & +52:23:33.1 & 3.5 & 3.6 & 92.68 & 3.08 & 458.4 & 24.7 & 21:09:21.85 & +52:23:40.9 & 2.4 & 2.6 & 616.5 & 36.4 & 100.0 & 142.8 & 70.9 & 48.0 & 5.08 & E \\
AMIGPS J212437+505842 & 21:24:37.19 & +50:58:42.8 & 2.9 & 3.1 & 93.32 & 0.4 & 317.4 & 16.6 & 21:24:35.76 & +50:59:01.8 & 2.6 & 2.5 & 619.9 & 33.8 & 100.0 & 328.3 & 0.0 & 129.8 & 12.60 & E \\
AMIGPS J210515+493959 & 21:05:15.32 & +49:39:59.7 & 3.3 & 3.5 & 90.23 & 1.72 & 309.2 & 16.5 & 21:05:15.70 & +49:40:12.3 & 2.3 & 2.5 & 398.6 & 23.2 & 100.0 & 112.5 & 71.8 & 131.0 & 4.01 & E \\
AMIGPS J210618+475108 & 21:06:18.59 & +47:51:08.2 & 2.4 & 2.6 & 89.0 & 0.38 & 233.6 & 12.0 & 21:06:19.13 & +47:51:09.3 & 1.8 & 2.1 & 226.0 & 12.2 & 100.0 & 0.0 & 0.0 & 0.0 & 3.36 & P \\
AMIGPS J211351+522729 & 21:13:51.78 & +52:27:29.9 & 10.0 & 10.0 & 93.2 & 2.63 & 218.7 & 17.8 & 21:13:49.23 & +52:27:54.6 & 5.8 & 5.6 & 319.2 & 33.9 & 136.9 & 140.9 & 82.1 & 127.0 & 1.59 & E \\
AMIGPS J211215+522451 & 21:12:15.05 & +52:24:51.0 & 18.7 & 18.7 & 93.0 & 2.77 & 212.6 & 28.0 & -- & -- & -- & -- & -- & -- & 188.3 & -- & -- & -- & -- & -- \\
AMIGPS J210217+470217 & 21:02:17.46 & +47:02:17.9 & 2.6 & 2.8 & 87.94 & 0.34 & 207.5 & 10.7 & 21:02:17.13 & +47:02:18.5 & 1.9 & 2.1 & 195.1 & 10.8 & 100.0 & 0.0 & 0.0 & 0.0 & 1.30 & P \\
AMIGPS J211221+521947 & 21:12:21.95 & +52:19:47.1 & 16.9 & 16.9 & 92.95 & 2.7 & 193.1 & 23.3 & -- & -- & -- & -- & -- & -- & 179.1 & -- & -- & -- & -- & -- \\
AMIGPS J211415+495345 & 21:14:15.31 & +49:53:45.2 & 3.5 & 3.6 & 91.39 & 0.81 & 154.9 & 8.4 & 21:14:15.17 & +49:53:45.3 & 2.2 & 2.5 & 152.3 & 9.3 & 100.0 & 29.3 & 0.0 & 5.4 & 1.40 & P \\
AMIGPS J210821+493644 & 21:08:21.89 & +49:36:44.2 & 4.1 & 4.2 & 90.53 & 1.31 & 111.8 & 6.2 & 21:08:21.83 & +49:36:42.4 & 2.4 & 2.6 & 104.5 & 6.9 & 100.0 & 0.0 & 0.0 & 0.0 & 1.56 & P \\
AMIGPS J211852+493700 & 21:18:52.75 & +49:37:00.2 & 3.6 & 3.7 & 91.71 & 0.08 & 107.1 & 5.8 & 21:18:52.73 & +49:36:55.2 & 2.2 & 2.4 & 93.6 & 5.9 & 100.0 & 0.0 & 0.0 & 0.0 & 2.13 & P \\
AMIGPS J210056+533133 & 21:00:56.25 & +53:31:33.0 & 4.7 & 4.8 & 92.68 & 4.78 & 96.1 & 5.6 & 21:00:56.17 & +53:31:34.6 & 2.6 & 2.8 & 83.5 & 6.0 & 100.0 & 0.0 & 0.0 & 0.0 & 1.50 & P \\
AMIGPS J210344+495157 & 21:03:44.32 & +49:51:57.9 & 5.2 & 5.3 & 90.22 & 2.04 & 87.3 & 5.2 & 21:03:44.24 & +49:51:58.8 & 2.9 & 3.1 & 87.5 & 6.5 & 100.0 & 51.9 & 0.0 & 42.4 & 0.88 & P \\
AMIGPS J210538+480716 & 21:05:38.23 & +48:07:16.9 & 4.1 & 4.2 & 89.13 & 0.64 & 85.9 & 4.8 & 21:05:37.94 & +48:07:13.8 & 2.4 & 2.6 & 77.8 & 5.2 & 100.0 & 0.0 & 0.0 & 0.0 & 1.73 & P \\
AMIGPS J205642+494004 & 20:56:42.45 & +49:40:04.3 & 4.8 & 5.0 & 89.32 & 2.76 & 74.9 & 4.4 & 20:56:42.21 & +49:40:00.6 & 2.7 & 2.9 & 70.0 & 5.1 & 100.0 & 52.8 & 0.0 & 47.8 & 1.22 & P \\
AMIGPS J204207+500351 & 20:42:07.24 & +50:03:51.9 & 6.1 & 6.2 & 88.14 & 4.84 & 71.5 & 4.5 & 20:42:07.65 & +50:03:53.6 & 3.1 & 3.4 & 63.0 & 5.4 & 106.4 & 0.0 & 0.0 & 0.0 & 2.55 & P \\
AMIGPS J211331+503908 & 21:13:31.59 & +50:39:08.1 & 4.8 & 4.9 & 91.86 & 1.42 & 71.0 & 4.1 & 21:13:31.36 & +50:39:08.6 & 2.5 & 2.8 & 60.2 & 4.4 & 100.0 & 0.0 & 0.0 & 0.0 & 1.06 & P \\
\hline
\end{tabular}

\end{table}
\end{landscape}
}

\subsection{G159.6+7.3}

As an example of a section of the survey containing mostly (if not all) extra-galactic sources, in Fig.~\ref{Fi:G159} we show a $6^{\circ}\times 6^{\circ}$ region of the extension covering the supernova remnant G159.6+7.3.  116 sources are detected in this field, all of which are compact.  This is a large ($\approx\,3^{\circ}\times 4^{\circ}$), faint remnant discovered in H$\alpha$ in the Virginia Tech Spectral Line Survey \citep{2010AJ....140.1163F}.  Although the supernova remnant is large in extent, it has sharp, filamentary structures at the rim which should be minimally affected by flux loss in the AMI map (see their figs. 3 and 4).  No emission is visible in the AMI map; the maximum noise inside the annulus containing the H$\alpha$ emission is 3.2\,mJy\,beam$^{-1}$ (the mean is 1.7\,mJy\,beam$^{-1}$) so we estimate a conservative $5\sigma$ upper limit on the filamentary emission at 15.72\,GHz of 16\,mJy\,beam$^{-1}$.  Assuming a typical spectral index of $\alpha = 0.5$, where $S \propto \nu^{-\alpha}$, this corresponds to a surface brightness limit of $\approx 10^{-21}$\,W\,m$^{-1}$\,Hz\,sr$^{-1}$ at 1\,GHz, with the caveat that smooth, extended emission will be resolved out.  A more detailed comparison would take into account the AMIGPS $uv$-sampling; this is available on request from the authors.

\afterpage{
\begin{figure}
    \includegraphics[bb=46 113 567 664, clip=, width=\linewidth]{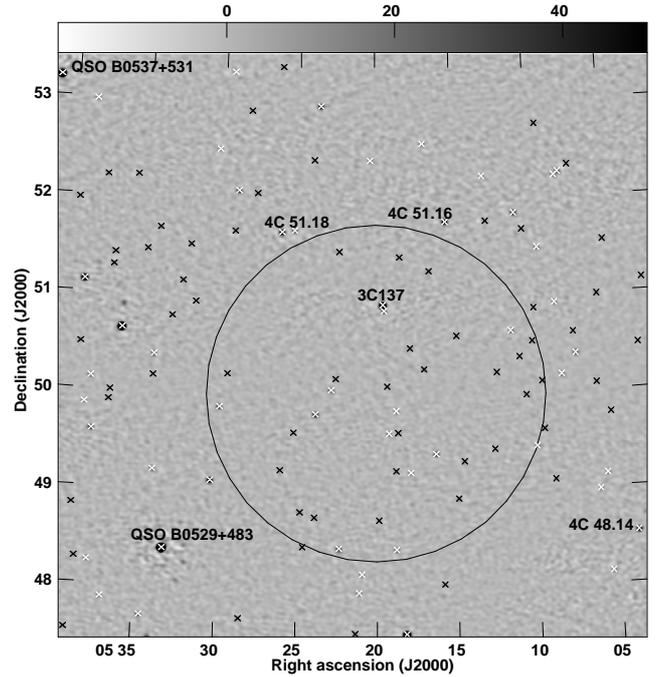}
    \caption{A $6^{\circ}\times 6^{\circ}$ region of the extension covering the supernova remnant G159.6+7.3.  The grey-scale is in mJy\,beam$^{-1}$ and has been truncated at $-20$ and $+50$ to show low-surface-brightness features.  Detected sources are marked with $\times$, and six of the brightest, known sources are labelled; they are all extra-galactic.  All of the sources in this field are compact and are likely to be extra-galactic.  The large circle marks the approximate outline of the supernova remnant, which is not detected.  The resolution of the map is 3 arcmin.}
    \label{Fi:G159}
\end{figure}
}

\section{Conclusions}

We present the second data release of the AMI Galactic Plane Survey, covering the Northern Galactic plane at $\delta \goa 20^{\circ}$ between $b \approx\,\pm 5^{\circ}$, corresponding to $53^{\circ} \lessapprox \ell \lessapprox 193^{\circ}$, plus high-latitude extensions to cover the Taurus and California giant molecular cloud regions and the supernova remnant G159.6+7.3. This extends the coverage of the survey from 868\,deg$^{2}$ in the first data release to 1777\,deg$^2$.  The resolution of the survey is $3$\,arcmin and the noise level is $\approx\,3$\,mJy\,beam$^{-1}$ away from bright sources.  This is the most sensitive and highest resolution large-area Galactic plane survey at cm-wave frequencies above 1.4\,GHz.

The pipeline used to produce the survey maps has been improved with respect to that used for DR-I, producing small improvements in positional accuracy (r.m.s.\ calibration errors are now 1.6\,arcsec and 1.9\,arcsec in RA and $\delta$ respectively) and flux calibration accuracy (flux densities of sources near the edge of strips are less likely to be biased downward).  The maps and catalogue of 6509 sources are available online from CDS and \url{http://www.astro.phy.cam.ac.uk/surveys/amigps}.

\section*{Acknowledgments}

We thank the staff of the Mullard Radio Astronomy Observatory for their invaluable assistance in the commissioning and operation of AMI, which is supported by Cambridge University and the Science and Technologies Facilities Council.  This work has made use of the distributed computation grid of the University of Cambridge (CAMGRID).  This research has made use of NASA's Astrophysics Data System Bibliographic Services and the facilities of the Canadian Astronomy Data Centre operated by the National Research Council of Canada with the support of the Canadian Space Agency. The research presented in this paper has used data from the Canadian Galactic Plane Survey, a Canadian project with international partners supported by the Natural Sciences and Engineering Research Council. YCP acknowledges support from a CCT/Cavendish Laboratory studentship and a Trinity College Junior Research Fellowship.  CR and TZJ acknowledge support from Science and Technology Facilities Council studentships.

\bsp \label{lastpage}

\end{document}